# Measured difference between $^{206}$Pb, $^{205}$Tl charge distributions and the proton $3s_{1/2}$ wave function

M. R. Anders$^a$, S. Shlomo$^{a,b}$ and I. Talmi$^b$

$^a$Cyclotron Institute, Texas A&M University, College Station, Texas 77840

$^b$The Weizmann Institute of Science, Rehovot 76100, Israel

Charge density difference between $^{206}$Pb and $^{205}$Tl, measured by elastic electron scattering, is very similar to the charge density due to a proton in a $3s_{1/2}$ orbit. We look for a potential well whose $3s_{1/2}$ wave function yields the measured data. We developed a novel method to obtain the potential directly from the density and its first and second derivatives. Fits to parametrized potentials were also carried out. The $3s_{1/2}$ wave functions of the potentials determined here, reproduce fairly well the experimental data within the quoted errors. To detect possible effects of short-range two-body correlations on the $3s_{1/2}$ shell model wave function, more accurate measurements are required.

1. Introduction

A very old problem is how the success of the nuclear shell model can be reconciled with the strong and short ranged interaction between free nucleons. It was realized that shell model wave functions are eigenstates of a *renormalized* nuclear Hamiltonian in which the interactions are rather tame. Thus, shell model wave functions, of independently moving nucleons, do not have short range correlations. The latter are imposed on the *real* wave functions by the strong short-range interaction between free nucleons (the bare interaction).

Still, there are indications that shell model wave functions have a certain reality. Many years ago, the difference between the charge distributions of $^{206}$Pb and $^{205}$Tl was measured by an ingenuous experiment [1, 2]. The difference of charge distributions which they determined is very similar to that due to a proton wave function in a $3s_{1/2}$ orbit. It is non-negative for all *r*, it has a clear maximum at the origin and two additional maxima. This result is in agreement with the simple shell model. The ½$^+$ ground state of $^{205}$Tl shows that the $3s_{1/2}$ orbit is the highest in the proton Z = 82 major shell. The experimental charge density near *r* = 0 is much lower than in Hartree-Fock calculations or by using a conventional Wood-Saxon potential. The authors attributed it to an admixture of a shell model wave function of a proton hole in a $2d_{3/2}$ state, coupled to a $J = 2$ state of two neutron holes.

This interpretation of the data seemed to be in contradiction with results of the nuclear many-body theory. In the latter, short-range two-nucleon correlations due to the interaction between free nucleons, play an important role. Indeed, the case described above was considered, among others, in a paper [3] where the admixtures due to two-body short range correlations were calculated to be rather high, up to 35%.

These estimates are based on measured cross sections of various reactions which are smaller than those calculated with shell model wave functions. Unlike the depletions considered there, the measured $^{206}$Pb - $^{205}$Tl charge difference is exactly equal to 1 proton charge. Any effect of short range correlations could only modify the *shape* of the difference between charge distributions.

Motivated by the striking results of the measured difference of the charge densities, we look in the present paper, for a potential whose proton $3s_{1/2}$ wave function can reproduce the measured difference. If such a potential exists, no effect of short range correlations is evident in the experimental data. This would not contradict the existence of those correlations. The latter cause "wounds" in shell model wave functions. If the wound occupies a small volume, it will not have a big effect on expectation values of "long range" operators. If such a potential is found, it could serve also as an additional constraint in the determination of a modern energy density functional (EDF) for more reliable prediction of properties of nuclei and nuclear matter [4,5]. For this purpose we have developed a new method to determine the single particle potential directly from the single particle matter density and its first and second derivatives [6].

## 2. Formalism

Consider the Schrodinger equation,

$$-\frac{\hbar^2}{2m}\Delta\psi + V\psi = E\psi, \tag{1}$$

where $V(\vec{r})$ is a real local and non-singular potential. From Eq. (1) follows that for a given single particle wave function $\psi(\vec{r})$, known for all $\vec{r}$, and given eigenvalue $E$, the corresponding single particle potential $V$ is uniquely determined by

$$V(\vec{r}) = E + \frac{\hbar^2}{2m}S(\vec{r}), \quad S(\vec{r}) = \frac{\Delta\psi(\vec{r})}{\psi(\vec{r})}. \tag{2}$$

For a non-singular $V$, $\Delta\psi(\vec{r}) = 0$ when $\psi(\vec{r}) = 0$. The relation for $[\psi(\vec{r})]^b$, where b is a positive integer, is given in Ref. [6]. Here, we consider the spherically symmetric case,

$$\psi_{nlj}(\vec{r}) = \frac{R_{nlj}(r)}{r} Y_{lj}, \tag{3}$$

where, $R_{nlj}(r)$ is the (one-dimensional) radial wave function for the orbit with principal number $n$, orbital angular momentum $l$ and total angular momentum $j$ and $Y_{lj}$ is the eigenfunction of the angular momenta $l$ and $j$. In the following, we limit the discussion to the proton $3s_{1/2}$ orbit. Therefore, the corresponding single particle potential for a nucleon is

$$V_{cen}(r) = E + \frac{\hbar^2}{2m}S(r) - V_{coul}(r), \quad S(r) = \frac{d^2R}{dr^2}\frac{1}{R(r)}, \tag{4}$$

where $V_{coul}(r)$ is the Coulomb potential. .

The single particle radial density $\rho(r)$ is related to the square of the radial wave function $R$ by

$$R^2(r) = 4\pi r^2 \rho(r). \tag{5}$$

From (5) it is possible to extract the wave function $R(r)$ and use Eq. (4) to deduce the corresponding single particle potential, but this leads to numerical complications. Therefore, we developed a method to determine the potential directly from the density and its first and second derivatives. Using Eq. (1) for the radial wave-function $R(r)$, we obtain the simple relations

$$S(r) = \frac{1}{2R^2}\left[\frac{d^2(R^2)}{dr^2} - \frac{1}{2}\left[\frac{1}{R}\frac{d(R^2)}{dr}\right]^2\right], \tag{6}$$

and

$$S(r) = \frac{1}{2\rho}\left[\frac{d^2\rho}{dr^2} + \frac{2}{r}\frac{d\rho}{dr} - \frac{1}{2\rho}\left(\frac{d\rho}{dr}\right)^2\right]. \tag{7}$$

A commonly used central nuclear potential is the Woods Saxon (WS) potential,

$$V(r) = V_0/[1 + exp((r - R_1)/a_0)], \tag{8}$$

where, $V_0$, $R_1$ and $a_0$ are the depth, half radius and diffuseness parameters of the potential, respectively. For the Coulomb potential we adopt the form obtained from a uniform charge distribution of radius $R$,

$$V_{coul}(r) = Ze^2 \begin{cases} (3 - r^2/R^2)/2R & r < R \\ 1/r & r > R \end{cases}, \tag{9}$$

with $R^2 = (5/3)\langle r^2 \rangle_c$, where $\langle r^2 \rangle_c$ is the charge mean square radius.

In elastic electron-nucleus scattering measurements the charge density distribution, $\rho_c(\vec{r})$, is determined by carrying out a phase shift analysis of the cross section [7]. In theoretical models the point proton density distribution, $\rho_p(\vec{r})$ is calculated. The difference between them is due to the finite size of the proton internal charge distribution. They are related by the convolution relation

$$\rho_c(\vec{r}) = \int \rho_p(\vec{r'}) \rho_{pfs}(\vec{r} - \vec{r'}) d^3\vec{r'}, \tag{10}$$

where $\rho_{pfs}(\vec{r})$ is the charge density distribution of the proton. The experimental elastic electron scattering data of a free proton can be well reproduced by the form

$$\rho_{pfs}(\vec{r}) = \frac{1}{8\pi a^3} e^{-r/a}, \tag{11}$$

where $a^2 = \frac{1}{12}r_{pfs}^2$ with $r_{pfs} = 0.85$ fm being the corresponding charge root mean square (rms) radius [7,8]. The Fourier transform for the charge density $\rho_{ch}(r)$, given by the convolution relation of Eq. (10), is given by

$$F_c(q) = F_{pfs}(q)F_p(q),  \quad (12)$$

where $F_c(q)$, $F_{pfs}(q)$ and $F_p(q)$, are the Fourier transfoms of $\rho_{ch}(\vec{r})$, $\rho_{pfs}(\vec{r})$ and $\rho_p(\vec{r})$, respectively. Eq. (12) can be used to determine the form factor $F_p(q)$, associated with the point proton density distribution $\rho_p(r)$. Then $\rho_p(r)$ can be obtained from $F_p(q)$ by the inverse Fourier transform and compared with theoretical predictions.

3. Results

In Fig. 1a we present (solid line) the experimental data [1,2] for the charge density difference,

$$\Delta\rho_c(r) = \rho_c(r;\ ^{206}\text{Pb}) - \rho_c(r;\ ^{205}\text{Tl}), \quad (13)$$

between the isotones $^{206}$Pb – $^{205}$Tl. It is normalized to a total charge of one proton which is replaced in the following, by 1. The dotted lines indicate the experimental uncertainty. The two nodes associated with the proton $3s_{1/2}$ orbit are clearly seen in the figure. The experimental values of the charge rms radii of $^{206}$Pb and $^{205}$Tl are 5.4897 and 5.4792 fm, respectively, leading to a value of 6.2822 fm for the charge rms radius of the proton $3s_{1/2}$ orbit. To assess the possible rearrangement effect (from $^{205}$Tl to $^{206}$Pb) on the charge rms radius of the 81 protons core in $^{206}$Pb, we assume that it increases by 0.005 fm, similar to the change between nuclei in this region [8]. The rearrangement effect is approximated (see Ref. [9]) by scaling the charge distribution of $^{205}$Tl so that the charge rms radius of the scaled density is equal to that of the 81 core protons in $^{206}$Pb. We thus obtain

$$\Delta\rho_{Rc}(r) = \rho_c(r;\ ^{206}\text{Pb}) - \alpha^3 \rho_c(\alpha r;\ ^{205}\text{Tl}), \tag{14}$$

where the scaling parameter $\alpha = 5.4792/(5.4792 + 0.005) = 0.9990$ is the ratio between the charge rms radius of $^{205}$Tl to that of the 81 core protons in $^{206}$Pb. The form of (14) guarantees that the integral of $\Delta\rho_{Rc}(r)$ is equal to 1. The result for the charge density is shown in Fig.1a (dashed line).

To extract the corresponding single particle potential, using Eqs. (4) and (6) or (7), we need the point proton distribution, $\Delta\rho_p(r)$. It is obtained by using Eqs. (11) and (12) to determine the point proton form factor, $F_p(q)$, and then obtain $\Delta\rho_p(r)$ by inverse Fourier transform. Using the relation (5) we determined the values of $R_p^2(r) = 4\pi r^2 \Delta\rho_p(r)$ as obtained from Figure 1a and shown by the solid line in Figure 1b. Similarly, $R_{Rp}^2(r) = 4\pi r^2 \Delta\rho_{Rp}(r)$ and the dashed line in Figure 1b, is obtained from the dashed line in Figure 1a. The dotted lines indicate the experimental uncertainty. We note that $R_p^2(r)$ as obtained from Figure 1a (solid line) is slightly negative at the first node (at ~ 2.6 fm) and above zero at the second node (r ~ 4.9 fm). In the vicinity of these minima, however, the experimental uncertainty in $R_p^2(r)$ is larger than its value. The magnitude of the difference between $\Delta\rho_p(r)$ and $\Delta\rho_{Rp}(r)$ is similar to that of the experimental uncertainty.

We tried to use the experimental $R_p^2(r)$ and $R_{Rp}^2(r)$ of Figure 1b, shown by the solid and dashed lines, respectively, to directly deduce the corresponding potentials by employing (4) and (6). The Coulomb potential of Eq. (9), with $R = 7.1$ fm, was adopted in the calculations. For non-singular potential $V$, the second derivative should vanish when $R(r) = 0$. As seen from Figure 1b, this condition is not fulfilled at the nodes of the experimental $R_p^2(r)$. Still, in the vicinity of these nodes, in the regions of $r = 2.0 - 3.0$ fm and $r = 4.5 - 5.5$ fm, the uncertainty in

$\Delta\rho_p(r)$ is larger than 50% of its value. Due to the large uncertainties in the experimental data, no reliable potential can be extracted.

Therefore, we considered several nuclear central potentials, each includes the Coulomb potential of Eq. (9), whose $3s_{1/2}$ wave functions yield charge distributions which fit best the measured one. The parameters of these potentials are obtained by least square fits of the calculated $R_p^2(r)$ to the corresponding experimental data. The fitted potential $V_F(r)$ is obtained by taking the values of the potential at the points $r = 3$, 6 and 9 fm as free parameters. The value of $V_F(0)$ is constrained to reproduce the experimental value of 7.25 MeV for the separation energy of the proton in the $3s_{1/2}$ orbit of $^{206}$Pb. The value of $V_F(12)$ is taken to be zero. The values of $V_F(r)$ between these points are determined by polynomial interpolation (solid line in Fig.2). From a fit to the experimental data of $R_p^2(r)$, solid line in Figure 1b, we obtained the values of $V_F(r)$ at $r = 0$, 3, 6, 9 fm as -55.98, -86.80, -30.46, -23.09 MeV, respectively. Similarly, the potential $V_{RF}(r)$ (dashed line) is obtained by a fit to $R_{Rp}^2(r)$ (dashed line of Figure 1b) with resulting values of -48.26, -95.89, -8.72, -23.29 MeV, for at $r = 0$, 3, 6, 9 fm, respectively.

The potential $V_{WSF}(r)$, dashed double dotted line, is obtained by fitting the Wood-Saxon potential, Eq. (8), obtaining the values of -167.95 MeV, -0.03 fm and 4.68 fm for $V_0$, $R_1$ and $a_0$, respectively. For comparison, we also show by the dashed-dotted line the Wood-Saxon Potential $V_{WS}(r)$ using the conventional values of

-62.712 MeV, 7.087 fm and 0.65 fm for $V_0$, $R_1$ and $a_0$, respectively.

In Figures 3a and 3b we compare the experimental results of $R_c^2(r) = 4\pi r^2 \Delta\rho_c(r)$ and $\Delta\rho_c(r)$ with the results from the proton $3s_{1/2}$ orbit, obtained from potentials shown in Figure 2. The experimental data is the region between the dotted lines. The results obtained using, $V_F(r)$, $V_{WSF}(r)$ and $V_{WS}(r)$ potentials are shown by the solid, dashed-double dotted and dashed-dotted curves, respectively. The results of the fitted $V_F(r)$ potential are in fair agreement with the experimental data, with $\chi^2/N = 1.15$. The results of the $V_{RF}$ potential (not shown in Fig.3) are also in fair agreement with data, with $\chi^2/N = 1.81$. The results of the fitted potential $V_{WSF}(r)$ are in reasonable agreement with data, with $\chi^2/N = 3.28$. The results of $V_{WS}$ are shown for comparison. The agreement with the data is much poorer, with $\chi^2/N = 8.85$. The amplitudes of the oscillations of $R_c^2(r)$ obtained from the conventional Wood-Saxon potential, $V_{WS}(r)$, are much larger than those of the experimental data for $r$ smaller than 5.0 fm and much smaller than the data for $r$ larger than 5.0 fm. As noted in Refs. [1,2], the calculated value of the charge density $\Delta\rho_c(r)$ at $r = 0$ obtained using the $V_{WS}(r)$ potential is larger than the experimental value by a factor two.

In Figures 4a, b and c we compare the charge density of the $1s_{1/2}$, $2s_{1/2}$ and $3s_{1/2}$ proton orbits, respectively, obtained from the fitted potential $V_F(r)$ (solid lines) with those obtained from the conventional Wood-Saxon potential $V_{WS}(r)$ (dashed-double dotted lines). The separation energies of the $1s_{1/2}$, $2s_{1/2}$ and $3s_{1/2}$ proton orbits are -47.09, -22.64 and -7.24 MeV for the $V_F(r)$ potential and -36.31, -24.46 and -8.00 MeV for the $V_{WS}(r)$ potential, respectively. The relatively large separation energy of the $1s_{1/2}$ proton orbit obtained for the $V_F(r)$ is close to the experimental data [10]. At $r = 0$ only the proton $s$ orbits contribute to the charge density, $\rho_c(r)$, in $^{206}$Pb. The calculated value of $\rho_c(0) = 0.060$ fm$^{-3}$ for the potential $V_F(r)$ is significantly smaller than the value of $\rho_c(0) = 0.073$ fm$^{-3}$ for a conventional Wood-Saxon potential, $V_{WS}(r)$. It is in good agreement with the experimental value of $\rho_c(0) = 0.063$ [1,2].

4. Conclusion

The difference between the charge distributions of $^{206}$Pb and $^{205}$Tl was measured many years ago. It offers a good opportunity to study possible effects of short range correlations on the shell model wave function of a proton in the $3s_{1/2}$ orbit. Effects of this kind were estimated by comparing measured cross sections of various reactions to those calculated using shell model wave functions. Usually, the measured values were lower than the calculated ones. The difference between the charge distributions considered here cannot be depleted. The integrated difference must be exactly equal to the charge difference between the two isotones, 1 proton charge. The effects of short range correlations in this case can only change the *shape* of the difference between the charge distributions.

The experimental values of that difference have features which are very similar to those due to the wave function of a proton in a $3s_{1/2}$ orbit. Within the experimental error bars, all its values are non-negative and there are two zero values for $r > 0$ which correspond to the two nodes of the $3s$ wave function $R_p(r)/r$. If the point proton distribution $\Delta\rho_p(r)$ is due to a $3s$ wave function, two more conditions should be satisfied, in addition to its having first and second derivatives for all $r$ values. At $r$ where $\Delta\rho_p(r) = 0$, also its first derivative should vanish. The condition that if $R_p(r) = 0$, also its second derivative must vanish at that $r$ leads to the other condition. Where the point proton distribution $\Delta\rho_p(r) = 0$, the corresponding expression in the square brackets on the right hand side of (7) should vanish. These conditions are necessary for deriving the single particle potential $V$, using Eqs. (4) and (6) or (7), from a parametrized point proton distribution $\Delta\rho_p(r)$ fitted to the experimental data. It is difficult to see whether these conditions are satisfied by the measured difference of charge distributions. The experimental accuracy is not sufficient, especially near the zero values.

We started by deriving and employing a new relation [6] between the potential $V$ and the single particle density and its first and second derivatives, Eqs. (4), (6) and (7). Around its minima, the experimental uncertainty in $\Delta\rho_c(r)$ is larger than its value. Hence, no reliable potential can be obtained.

In view of this situation, we tried to construct nuclear single particle potentials $V$ whose proton $3s_{1/2}$ orbit in $^{206}$Pb yield charge distributions which best fit the electron scattering data. We found several potentials which yield fair fits to the data (Fig. 3). The fair agreement with fitted potentials may be an indication that effects of short range correlations on charge distributions due to shell model wave functions are not significant. More accurate experimental data for $\Delta\rho_c(r)$ with uncertainty smaller by a factor of two or more may answer the question how well can the data be reproduced by a calculated $3s_{1/2}$ wave function.

Acknowledgements

S. S. would like thank The Weizmann Institute of Science for the kind hospitality. This work was supported in part by US Department of Energy under Grant No. DOE-FG03-93ER40773.

# FIGURE CAPTIONS

Fig.1(a) The experimental difference, $\Delta\rho_c(r)$ between $^{206}$Pb and $^{205}$Tl charge distributions (solid line). The dashed line is for $\Delta\rho_{Rc}(r)$, the data after rearrangement correction. The dotted lines indicate the experimental uncertainty. (b) Similar to (a) for $R_P^2(r) = 4\pi r^2 \Delta\rho_p(r)$ where $\Delta\rho_p(r)$ is derived from the experimental $\Delta\rho_c(r)$. The dashed line is for $R_{Rp}^2(r)$ related to $\Delta\rho_{Rp}(r)$ similarly obtained from $\Delta\rho_{Rc}(r)$.

Fig.2 Potentials fitted to data in Fig.1b. The $V_F(r)$ potential (solid line), the $V_{FR}(r)$ version including rearrangement (dashed line) and the fitted $V_{FWS}(r)$ potential (double dotted-dashed line). Also shown is the conventional Wood-Saxon $V_{WS}(r)$ potential (dashed-dotted line).

Fig.3 Experimental values of $R_c^2(r) = 4\pi r^2 \Delta\rho_c(r)$ (a) and $\Delta\rho_c(r)$ (b) plotted between dotted lines of error limits. They are compared to calculated charge distributions due to the $3s_{1/2}$ wave functions of the fitted $V_F(r)$ potential (solid lines), the fitted Wood-Saxon $V_{FWS}(r)$ potential (double dotted-dashed lines) and the conventional $V_{WS}(r)$ potential (dashed-dotted lines.

Fig.4 Calculated charge densities of a proton in the $1s_{1/2}$ (a), $2s_{1/2}$ (b) and $3s_{1/2}$ (c) orbits in the $V_F(r)$ potential (solid lines) and the conventional $V_{WS}(r)$ potential (double dotted-dashed lines).

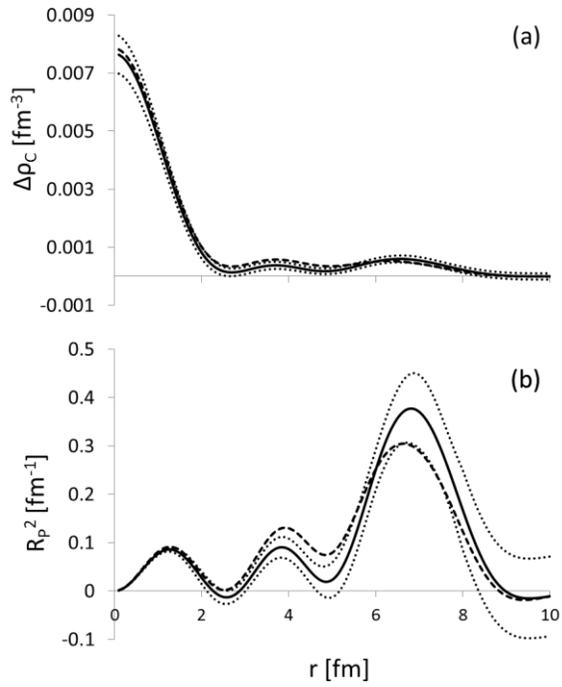

Fig.1(a) The experimental difference, $\Delta\rho_c(r)$ between $^{206}$Pb and $^{205}$Tl charge distributions (solid line). The dashed line is for $\Delta\rho_{Rc}(r)$, the data after rearrangement correction. The dotted lines indicate the experimental uncertainty. (b) Similar to (a) for $R_P^2(r) = 4\pi r^2 \Delta\rho_p(r)$ where $\Delta\rho_p(r)$ is derived from the experimental $\Delta\rho_c(r)$. The dashed line is for $R_{Rp}^2(r)$ related to $\Delta\rho_{Rp}(r)$ similarly obtained from $\Delta\rho_{Rc}(r)$.

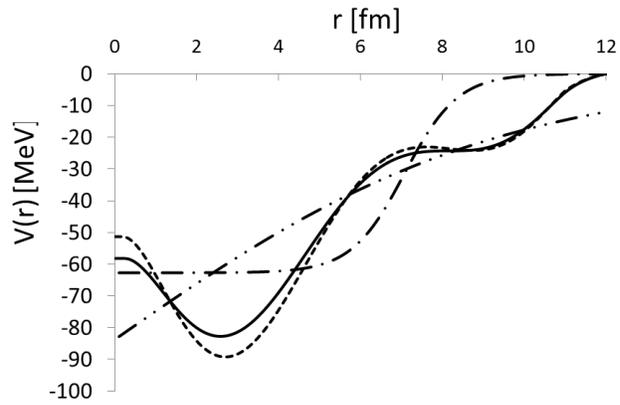

Fig.2 Potentials fitted to data in Fig.1b. The $V_F(r)$ potential (solid line), the $V_{FR}(r)$ version including rearrangement (dashed line) and the fitted $V_{FWS}(r)$ potential (double dotted-dashed line). Also shown is the conventional Wood-Saxon $V_{WS}(r)$ potential (dashed-dotted line).

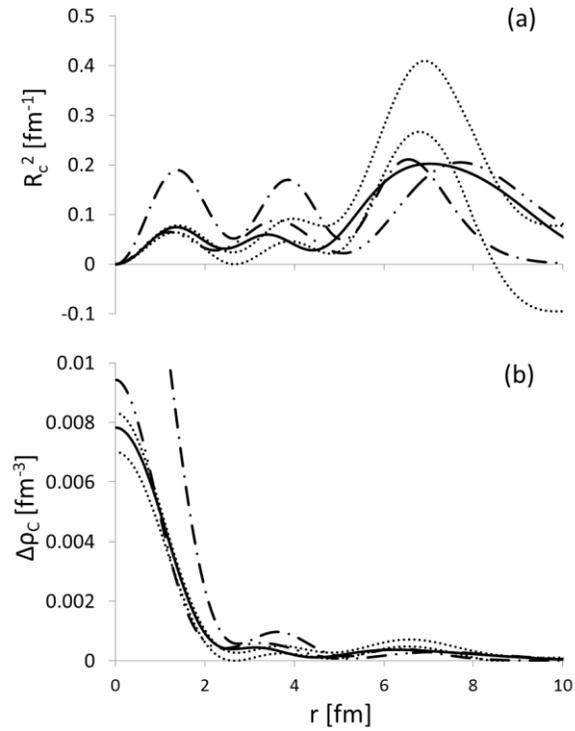

Fig.3 Experimental values of $R_c^2(r) = 4\pi r^2 \Delta\rho_c(r)$ (a) and $\Delta\rho_c(r)$ (b) plotted between dotted lines of error limits. They are compared to calculated charge distributions due to the $3s_{1/2}$ wave functions of the fitted $V_F(r)$ potential (solid lines), the fitted Wood-Saxon $V_{FWS}(r)$ potential (double dotted-dashed lines) and the conventional $V_{WS}(r)$ potential (dashed-dotted lines.

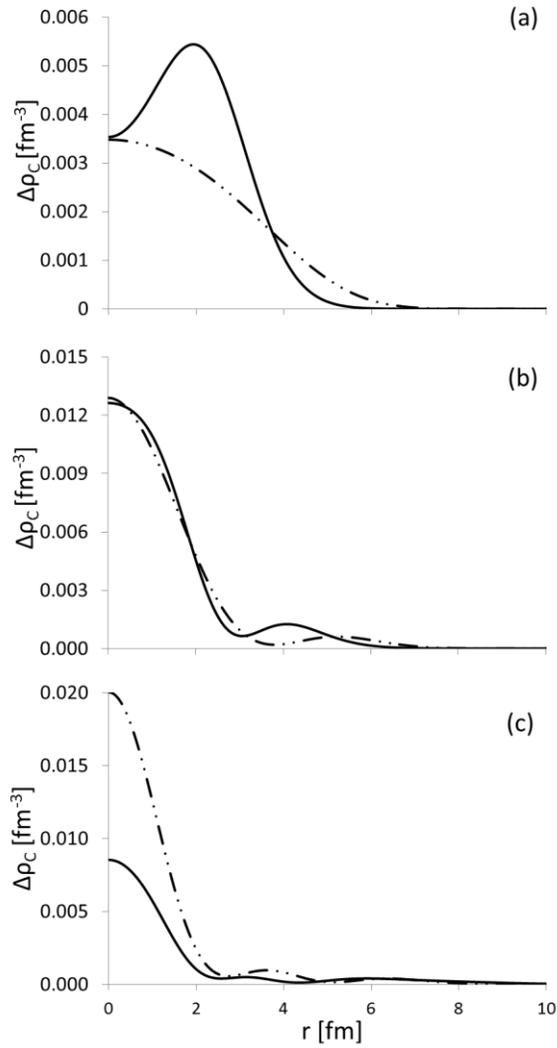

Fig.4 Calculated charge densities of a proton in the $1s_{1/2}$ (a), $2s_{1/2}$ (b) and $3s_{1/2}$ (c) orbits in the $V_F(r)$ potential (solid lines) and the conventional $V_{WS}(r)$ potential (double dotted-dashed lines).